\documentclass[12pt,a4paper]{article}
\usepackage[applemac]{inputenc}
\usepackage{amssymb}
\usepackage{amsmath}
\usepackage{amsthm}
\usepackage{amsopn}
\usepackage[pdftex,bookmarks,colorlinks,breaklinks]{hyperref}  
\hypersetup{linkcolor=red,citecolor=blue,filecolor=dullmagenta,urlcolor=darkblue}

\newcommand{\CC}{\mathbb{C}}

\newcommand{\KK}{\mathbb{K}}

\newcommand{\NN}{\mathbb{N}}

\newcommand{\RR}{\mathbb{R}}

\newcommand{\frA}{\mathfrak{A}}

\newcommand{\frAr}{\mathfrak{A}(\rr)}

\newcommand{\frb}{\mathfrak{B}}

\newcommand{\frsr}{\mathfrak{S}(\rr)}

\newcommand{\kaa}{\mathcal{A}}
\newcommand{\kbb}{\mathcal{B}}

\newcommand{\kcc}{\mathcal{C}}

\newcommand{\cor}{\mathcal{C}\mathcal{O}\mathcal{N}(\rr)}
\newcommand{\kD}{\mathcal{D}}

\newcommand{\kf}{\mathcal{F}}

\newcommand{\kh}{\mathcal{H}}
\newcommand{\kI}{\mathcal{I}}
\newcommand{\kj}{\mathcal{J}}

\newcommand{\kL}{\mathcal{L}}
\newcommand{\mm}{\mathcal{M}}

\newcommand{\kO}{\mathcal{O}}
\newcommand{\kP}{\mathcal{P}}
\newcommand{\kQ}{\mathcal{Q}}
\newcommand{\kR}{\mathcal{R}}
\newcommand{\kS}{\mathcal{S}}
\newcommand{\ksr}{\kS(\rr)}
\newcommand{\ksm}{\kS(\mm)}

\newcommand{\ksa}{\kS(\kaa)}

\newcommand{\kT}{\mathcal{T}}

\newcommand{\gb}{\beta}
\newcommand{\gd}{\delta}
\newcommand{\eps}{\varepsilon}
\newcommand{\gga}{\gamma}
\newcommand{\gG}{\Gamma}

\newcommand{\gk}{\kappa}
\newcommand{\gl}{\lambda}
\newcommand{\go}{\omega}
\newcommand{\gO}{\Omega}
\newcommand{\gf}{\varphi}
\newcommand{\gfa}{\gf_{\kaa}}

\newcommand{\nua}{\nu_{\kaa}}
\newcommand{\nub}{\nu_{\kbb}}

\newcommand{\gF}{\Phi}
\newcommand{\gr}{\varrho}
\newcommand{\gs}{\sigma}
\newcommand{\gS}{\Sigma}
\newcommand{\gt}{\tau}

\newcommand{\tm}{\subseteq}

\newcommand{\∞}{\infty}

\newtheorem{theorem}{Theorem}[section]
\newtheorem{definition}{Definition}[section]
\newtheorem{proposition}{Proposition}[section]
\newtheorem{lemma}{Lemma}[section]
\newtheorem{corollary}{Corollary}[section]
\newtheorem{remark}{Remark}[section]
\newtheorem{example}{Example}[section]

\newcommand{\por}{\kP_{0}(\kR)}
\newcommand{\poa}{\kP_{0}(\kaa)}

\newcommand{\pom}{\kP_{0}(\mm)}

\newcommand{\pr}{\kP(\kR)}

\newcommand{\puh}{\kP_{1}(\lh)}
\newcommand{\puc}{\kP_{1}(\kL(\CC^3))}
\newcommand{\pa}{\kP(\kaa)}
\newcommand{\pb}{\kP(\kbb)}

\newcommand{\pmm}{\kP(\mm)}

\newcommand{\orr}{\kO(\kR)}
\newcommand{\omm}{\kO(\mm)}

\newcommand{\oaa}{\kO(\kaa)}

\newcommand{\qr}{\mathcal{Q}(\mathcal{R})}

\newcommand{\qpr}{\mathcal{Q}_{P}(\mathcal{R})}
\newcommand{\qpa}{\mathcal{Q}_{P}(\mathcal{A})}
\newcommand{\qpb}{\mathcal{Q}_{P}(\mathcal{B})}

\newcommand{\qa}{\mathcal{Q}(\mathcal{A})}
\newcommand{\qb}{\mathcal{Q}(\mathcal{B})}

\newcommand{\qmm}{\mathcal{Q}(\mm)}

\newcommand{\lh}{\mathcal{L}(\mathcal{H})}

\newcommand{\all}{\forall}
\newcommand{\ex}{\exists}
\newcommand{\rr}{\kR}

\newcommand{\hr}{\kR_{sa}}
\newcommand{\el}{E_{\gl}}

\newcommand{\emm}{E_{\mu}}

\newcommand{\eal}{E^{A}_{\gl}}

\newcommand{\eamu}{E^{A}_{\mu}}
\newcommand{\ebl}{E^{B}_{\gl}}

\newcommand{\eakl}{E^{A_{\gk}}_{\gl}}
\newcommand{\ea}{E^{A}}
\newcommand{\eak}{E^{A_{\gk}}}
\newcommand{\eb}{E^{B}}
\newcommand{\ebmu}{E^{B}_{\mu}}
\newcommand{\ec}{E^{C}}
\newcommand{\ecl}{E^{C}_{\gl}}

\newcommand{\we}{\wedge}
\newcommand{\We}{\bigwedge}
\newcommand{\Ve}{\bigvee}

\newcommand{\tto}{\mapsto}
\newcommand{\lra}{\Longrightarrow}
\newcommand{\llra}{\Longleftrightarrow}

\newcommand{\dm}{\kD(\mm)}
\newcommand{\dr}{\kD(\rr)}

\newcommand{\kcb}{\gr^{\kcc}_{\kbb}}
\newcommand{\kca}{\gr^{\kcc}_{\kaa}}
\newcommand{\kba}{\gr^{\kbb}_{\kaa}}

\newcommand{\irr}{\in \RR}

\newcommand{\lir}{\gl \in \RR}

\newcommand{\ikk}{\in \KK}
\newcommand{\kik}{k \in \KK}

\newcommand{\urb}{\overset{-1}}

\newcommand{\pcx}{P_{\CC x}}

\begin{document}

\title{\Huge{Observables \\ IV : The Presheaf Perspective}}

\author{Hans F.\ de Groote\footnote{degroote@math.uni-frankfurt.de;
FB Mathematik, J.W.Goethe-Universität Frankfurt a.\ M.}}

\titlepage
\maketitle
\tableofcontents
\begin{abstract}
In this fourth of our series of papers on observables we show that one can associate to each von Neumann algebra $R$ a pair of isomorphic presheaves, the upper presheaf $O^{+}_{R}$and the lower presheaf $O^{-}_{R}$, on the category of abelian von Neumann subalgebras of $R$. Each $A \in R_{sa}$ induces a global section of $O^{+}_{R}$ and of $O^{-}_{R}$ respectively. We call them \emph {contextual observables}. But we show that, in general, not every global section of these presheaves arises in this way. Moreover, we discuss states of a von Neumann algebra in the presheaf context. 
\end{abstract}
\pagebreak  
\begin{center}
    {\large Für Karin}
\end{center}

\section{Introduction}
\label{in}

In this fourth paper of our series on ``observables'' (\cite{deg3}, \cite{deg4}, \cite {deg5}) we generalize the notion of a (bounded) quantum observable in the sense of contextuality. The main results where already announced in our overview article \cite{deg2a}. 
~\\
The central idea in this article is the notion of restriction of operators from a von Neumann algebra $\rr$ in $\lh$ (for an arbitrary Hilbert space $\kh$) to a von Neumann subalgebra $\mm$ of $\rr$. This restriction of operators turns out to be a far reaching generalization of the notion of central support (or central carrier) of a projection. If $P \in \pr$, we call $s_{\mm}(P):= \We \{ Q \in \pmm \mid Q\geq P \}$ the $\mm$-support of $P$. We can generalize this definition to selfadjoint operators $A \in \hr$ by setting 
\begin{displaymath}
\gr_{\mm}A := \We \{ B \in \mm_{sa} \mid A \leq_{s} B \},  
\end{displaymath}
where $\leq_{s}$ denotes the spectral order on $\hr$ (\cite {deg2, [O]}), called the restriction of $A$ to $\mm$. One can prove directly from this definition that $\gr_{\mm}A$ is a selfadjoint element of $\mm$: If $B \in \mm$ such that $A \leq B$, then $ (\min sp (A)) I \leq_{s} B$, hence the bounded completeness of the spectral order implies that $\gr_{\mm}A \in \mm$. However, in order to gain a deeper insight into the restriction process, we show how the restriction of $A$ is obtained in a natural way from the observable function of $A$. For the convenience of the reader we will revise the basic definitions and results on observable functions. \\
Let $\rr$ be a von Neumann algebra acting on a Hilbert space $\kh$ and let $A \in \rr$ be selfadjoint. Moreover, let $\dr$ be the set of all dual ideals of the projection lattice $\pr$ of $\rr$ and let $\qr \tm \dr$ be the set of all maximal dual ideals (the ``quasipoints ''). $\qr$ is called the Stone spectrum of $\rr$. It is, equipped with the topology that is generated by the sets $\qpr:= \{ \frb \in \qr \mid P \in \frb \} \ (0 \neq P \in \pr) $, a zero - dimensional Hausdorff space, for which the clopen\footnote {We use the shortcut ``clopen '' for ``closed and open ''.} sets $\qpr$ form a base. The space $\dr$ bears the analogously defined topology. But note that $\dr$, equipped with this topology, is (in general) not a Hausdorff space. If $\rr$ is abelian, $\qr$ is homeomorphic to the Gelfand spectrum $\gO (\rr)$ of $\rr$ (\cite {deg3}). If $A \in \hr$ and if $E = (\el)_{\lir}$ is the spectral family corresponding to $A$, the function $f_{A} : \dr \to \RR$, defined by 
\begin{displaymath}
f_{A}(\kj) := \inf \{ \lir \mid \el \in \kj \}, 
\end{displaymath}  
is called the observable function of $A$. Its restriction to the Stone spectrum $\qr$ of $\rr$ is \emph{continuous}, whereas $f_{A}:\dr \to \RR$ is, in general, not. We note further that, if $\rr$ is abelian, $f_{A}:\qr \to \RR$ is (up to the homeomorphism $\qr  \cong \gO (\rr) $ mentioned before) the Gelfand transform of $A$. \\
We can characterise observable functions in an abstract way (\cite {deg4}): 
\begin{theorem}\label {in1} 
    $f : \dr \to \RR$ is an observable function if and only if the
    following two properties hold for $f$:
    \begin{enumerate}
       \item  [(i)]  $\all \ \kj \in \dr  : \ f(\kj) = \inf \{ f(H_{P}) |
       \ P \in \kj \}$,
       
       \item  [(ii)] $f(\bigcap_{j \in J}\kj_{j}) = \sup_{j \in J}f(\kj_{j})$
	for all families $(\kj_{j})_{j \in J}$ in $\dr$.
    \end{enumerate}
\end{theorem}
Here $H_{P}$ denotes the principle dual ideal defined by $P$: $H_{P}:= \{ Q \in \pr \mid P \leq Q \}$. An observable function $f_{A}$ on $\dr$ induces a function $r_{A}: \por \to \RR$, given by $r_{A}(P):= f_{A}(H_{P})$. This function is bounded and has the property 
\begin{displaymath}
r_{A}(\Ve_{\kik}P_{k}) = \sup_{\kik}r_{A}(P_{k})
\end{displaymath}
for all families $ (P_{k})_{\kik}$ in $\por$. Therefore, it is called \emph{completely increasing}. Conversely, every bounded completely increasing function $r: \por \to \RR$ comes from a unique selfadjoint operator $A \in \hr$:
\begin{displaymath}
\ex! \ A \in \hr: \ r=r_{A}.
\end{displaymath}
If $\mm$ is a von Neumann subalgebra of the von Neumann algebra $\rr$, and if $r: \por \to \RR$ is a bounded completely increasing function, $r$ can be restricted to a bounded function $\gr_{\mm}r : \pom \to \RR$. $\gr_{\mm}r$ is obviously completely increasing. It is shown that the selfadjoint operator corresponding to $\gr_{\mm}r$ is $\gr_{\mm}A$, where $A$ is the selfadjoint operator corresponding to $r$. It is in this sense that $\gr_{\mm}A$ is a restriction of $A$. \\
~\\
The foregoing definition of restricting selfadjoint operators from $\rr$ to $\mm$ has an equally natural counterpart: 
\begin{displaymath}
\gs_{\mm}A := \Ve \{ B \in \mm_{sa} \mid A \geq_{s} B \}. 
\end{displaymath}
We show in the sequel that this type of restriction is induced by the \emph{mirrored observable function} $g_{A}$ of $A$. Mirrored observable functions were introduced by A. D\"{o}ring in \cite {doe1b} and called \emph{antonymous} functions. $g_{A}$ is defined on $\dr$ by 
\begin{displaymath}
\all \ \kj \in \dr : \ g_{A}(\kj) := \sup \{ \lir \mid I - \el \in \kj \}, 
\end{displaymath}
where, as usual, $ (\el) _{\lir}$ denotes the spectral family corresponding to $A$. There is a simple relation between the observable function and the mirrored observable function of $A$: 
\begin{displaymath}
g_{A} = -f_{-A}.
\end{displaymath}
Hence, if $\orr$ denotes the set of all observable functions, the set of all mirrored observable functions is just $-\orr$. Therefore, we prefer the name ``mirrored observable function '' instead of ``antonymous function ''. It follows immediately from the foregoing relation between observable and mirrored observable functions that a theorem analogous to theorem \ref {in1} is true: 

\begin{theorem}\label {in2} 
    $g : \dr \to \RR$ is a mirrored observable function if and only if the
    following two properties hold for $g$:
    \begin{enumerate}
       \item  [(i)]  $\all \ \kj \in \dr  : \ g(\kj) = \sup \{ g(H_{P}) |
       \ P \in \kj \}$,
       
       \item  [(ii)] $g(\bigcap_{j \in J}\kj_{j}) = \inf_{j \in J}g(\kj_{j})$
	for all families $(\kj_{j})_{j \in J}$ in $\dr$.
    \end{enumerate}
\end{theorem}
The mirrored observable function $g_{A}$ induces a bounded function $s_{A}:\por \to \RR$, defined by $s_{A}(P):= g_{A}(H_{P})$, which is \emph{completely decreasing}:
\begin{displaymath}
s_{A}(\Ve_{\kik}P_{k})=\inf_{\kik}s_{A}(P_{k})
\end{displaymath}
for all families $ (P_{k})_{\kik}$ in $\por$. Using theorem \ref {in2}, one can easily  show that every bounded completely decreasing function $s: \por \to \RR$ is induced by a unique operator $A \in \hr$: $s = s_{A}$. \\
If $\mm$ is a von Neumann subalgebra of $\rr$ and $A \in \hr$ is an operator with corresponding completely decreasing function $s_{A}: \por \to \RR$, we prove that $\gs_{\mm}A$ is the selfadjoint operator corresponding to the restriction $s_{A}|_{\pom}$ of $s_{A}$ to $\pom$. \\
~\\
In section \ref {ls} we present a new approach to the spectral order on $\hr$ via observable functions. Section \ref {res} contains a detailed discussion of the restriction processes defined above. \\
In section \ref {ulp} we define the upper and lower observable presheaf of a von Neumann algebra $\rr$. These are presheaves on the category $\frAr$ of abelian von Neumann subalgebras of $\rr$ (the \emph {context category}), hence presheaves in the sense of Topos theory\footnote {Of course $\frAr$ can be regarded here simply as a semi-lattice.}. The restrictions of the upper (lower) observable presheaf are defined by restricting completely increasing (decreasing) functions corresponding to selfadjoint operators. It is shown that the upper observable presheaf is isomorphic to the lower one. \\
In section \ref {uni} we present a possible unification of upper and lower observable presheaves to a presheaf of linear spaces and linear maps. This unification is mathematical natural. However, we don't have an operator - theoretical interpretation for it.\\
In section \ref {gs} we discuss global sections of observable presheaves. Here, the issue is not their existence (every selfadjoint operator gives rise to a global section of the observable presheaf\footnote {We discuss in this section only the upper observable presheaf. The same results hold for the lower observable presheaf.}) but the question whether every global section of the observable presheaf is induced by a single selfadjoint operator. Of course, this is not true when the von Neumann algebra contains a direct summand of type $\rm {I} _{2}$, but we give an example that is different from this situation. Thus the global sections of observable presheaves form a larger class than the usual observables. We call the global sections of the upper (lower) observable presheaf \emph {upper (lower) contextual observables}. These constructions play an essential r\^{o}le in the remarkable articles of A. D\"{o}ring and C.J. Isham (\cite {id1, id2}) on a Topos - theoretical formulation of physical theories. \\ In section \ref {st} we consider states of a von Neumann algebra $\rr$ from the presheaf perspective. We show that each state of $\rr$ induces a global section of the \emph{state presheaf } $\kS_{\rr}$. $\kS_{\rr}$ is a presheaf on the context category $\frAr$ of abelian von Neumann subalgebras of $\rr$, which assigns to each $\kaa \in \frAr$ the space $\kS (\kaa) $ of all states of $\kaa$; the restrictions are the ordinary restrictions of functions. We show that, if $\rr$ contains a direct summand of type $\rm {I} _{2}$, not every global section of $\kS_{\rr}$ is induced by a state of $\rr$. But, contrary to the observable presheaves, this is the only exception. It follows from the generalization of Gleason's theorem, due to Christensen, Yeadon et al. (\cite{mae}), that each global section of $\kS_{\rr}$ is induced by a state of $\rr$, provided that $\rr$ does not contain a summand of type $\rm {I} _{2}$. More precisely, we show that this generalization of Gleason's theorem is equivalent to the property that every global section of $\kS_{\rr}$ is induced by a state of $\rr$.\\ 
~\\ 
We emphasise that our results, except those of section \ref {st}, hold not only for von Neumann algebras but also for arbitrary complete orthomodular lattices: simply replace ``operators'' by ``bounded spectral families'' and ``abelian von Neumann subalgebras'' by ``complete Boolean sublattices''. This has the interesting consequence that the whole theory is applicable to the lattice of causally closed subsets (\cite{cas}) of an arbitrary spacetime. Whether this has consequences for general relativity should be investigated.

\section{A Canonical Lattice Structure on $\hr$}
\label{ls}

We have seen that selfadjoint operators $A \in \hr$ can be encoded in 
completely increasing functions $f : \por \to \RR$. In the sequel we
will make no distinction in the notation of completely increasing functions
and observable functions. Let $\orr$ be the set of observable functions.
Depending on the context these are either functions $f : \por \to \RR$
or $f : \dr \to \RR$ or $f : \qr \to \RR$. \\
~\\
There is a canonical partial order on $\hr$:

\begin{definition}\label{ls1}
    Let $A, B$ be selfadjoint elements of the von Neumann algebra
    $\rr$ and let $f_{A}, f_{B} : \por \to \RR$ be the observable functions
    corresponding to $A$ and $B$ respectively. Then we define
    \[
	A ≤_{s} B \quad \text{if and only if} \quad  f_{A} ≤ f_{B}
    \] 
    with respect to the pointwise defined ordering of real valued
    functions.
\end{definition}
It is easy to see that for $f, g \in \orr$ the relation $f ≤ g$ does
not depend whether we view these functions as being defined on $\por, 
\dr$ or $\qr$.

\begin{proposition}\label{ls2}
    Let $A, B \in \hr$ with spectral families $\ea$ and $\eb$
    respectively. Then
    \[
	A ≤_{s} B \ \llra \ \all \ \gl \in \RR : \ \ebl ≤ \eal.
    \]
\end{proposition}
\emph{Proof:} Let $A ≤_{s} B$. By definition we have 
\[
    A ≤_{s} B \ \llra \ \all \ P \in \por : \ \inf \{ \mu | P ≤
    \ea_{\mu} \} ≤ \inf \{ \mu | P ≤ \eb_{\mu} \}. 	 
\]
Let $\ebl \neq 0$. Then $\inf \{ \mu | \ebl ≤ \eb_{\mu} \} ≤ \gl$ and
therefore 
\[ 
     \gl_{0} :=\inf \{ \mu | \ebl ≤ \ea_{\mu} \} ≤ \gl.
\]
But then $\ebl ≤ \ea_{\gl_{0}} ≤ \eal$. Conversely, if $\ebl ≤ \eal$
for all $\gl \in \RR$, then obviously
\[
    f_{A}(P) = \inf \{ \gl | P ≤ \eal \} ≤ \inf \{ \gl | P ≤ \ebl \} =
    f_{B}(P) 
\]
for all $P \in \por$. \ \ $\Box$ \\ 
~\\
We call $≤_{s}$ the {\bf spectral order on $\hr$}. It is obvious that
$≤_{s}$ defines a partial order on $\hr$. \\ The spectral order was 
defined and studied by M.P. Olson in \cite{[O]} and independently in
\cite{deg2}. Originally the spectral order has been defined directly
by means of the spectral families corresponding to the selfadjoint
operators: $A ≤_{s} B$ if and only if $\ebl ≤ \eal$ for all $\gl \in
\RR$.\\
\emph{We think, however, that its most natural definition occurs here 
in connection with observable functions.}\\
The lattice operations were defined as follows:\\
Let $(A_{\gk})_{\gk \in \KK}$ be an arbitrary family in $\hr$ and let 
$\eak$ be the spectral family corresponding to $A_{\gk}$. Then 
\[
    \gl \tto \We_{\gk}\eakl \quad \text{and} \quad \gl \tto \We_{\mu >
    \gl}\Ve_{\gk}\eak_{\mu}
\]
are spectral families and the first of them defines the join
$\Ve_{\gk}A_{\gk}$, the second the meet $\We_{\gk}A_{\gk}$ of the
family $(A_{\gk})_{\gk \in \KK}$. With these operations of join and
meet $\hr$ is a boundedly complete lattice.

\section{Restrictions}
\label{res}

The abstract characterization of (quantum) observable functions leads 
to a natural definition of restricting selfadjoint elements of a von
Neumann algebra $\rr$ to a subalgebra $\mm$. Again we denote a
completely increasing function on $\por$ and the corresponding observable
function (on $\qr$ or $\dr$) by the same letter and speak simply of an
observable function. Obviously we have
\begin{remark}\label{PS1}
    Let $\mm$ be a von Neumann subalgebra of a von Neumann algebra
    $\rr$ and let $f : \por \to \RR$ be an observable function. Then
    the restriction
    \[
	\gr_{\mm}f := f_{\mid \pom}
    \]   
is an observable function for $\mm$. It is called the restriction of
$f$ to $\mm$.    
\end{remark}
This definition is absolutely natural. However, if $A$ is a
selfadjoint operator in $\rr$ then the observable function $f_{A} :
\por \to \RR$ corresponding to $A$ is a rather abstract encoding
of $A$. So before we proceed, we will describe the restriction map
\[
    \begin{array}{ccc}
	\gr_{\mm} : \orr & \to & \omm  \\
	f_{A} & \tto & \gr_{\mm}f_{A}
    \end{array}
\] 
in terms of spectral families. \\

To this end we define
\begin{definition}\label{PS2}
    Let $\kf$ be a filterbasis in $\por$. Then
    \[
	C_{\rr}(\kf) := \{ Q \in \por \ | \ \ex \ P \in \kf : \ P ≤ Q 
	\}
    \]
    is called the cone over $\kf$ in $\rr$.
\end{definition}
Clearly $C_{\rr}(\kf)$ is a dual ideal and it is easy to see that it
is the \emph{smallest dual ideal that contains $\kf$.} A dual ideal
$\kI \in \dm$ is, in particular, a filterbasis in $\pr$, so
$C_{\rr}(\kI)$ is well defined. 

\begin{proposition}\label{PS3}
    Let $f \in \orr$. Then 
    \[
	(\gr_{\mm}f)(\kI) = f(C_{\rr}(\kI))
    \]
    for all $\kI \in \dm$.
\end{proposition}
\emph{Proof:} From $f (\kj) = \inf_{P \in \kj}f (P) $, the definition of the cone and the fact that $f$ is increasing on $\por$ we obtain
\[
    f(C_{\rr}(\kI)) = \inf \{ f(Q) \ | \ Q \in C_{\rr}(\kI) \} = \inf \{
    f(P) \ | \ P \in \kI \}. \ \ \Box 
\]

\begin{definition}\label{PS4}
    For a projection $Q$ in $\rr$ let
    \[
	c_{\mm}(Q) := \Ve \{ P \in \pmm \ | \ P ≤ Q \} \quad
	\text{and} \quad s_{\mm}(Q) := \We \{ P \in \pmm \ | \ P ≥ Q \}.
    \]
    $c_{\mm}(Q)$ is called the {\bf $\mm$-core}, $s_{\mm}(Q)$ the
    {\bf $\mm$-support} of $Q$.
\end{definition}
The $\mm$-support is a natural generalization of the notion of central
support which is the $\mm$-support if $\mm$ is the center of $\rr$. Note
that if $Q \notin \mm$ then $c_{\mm}(Q) < Q < s_{\mm}(Q)$.
The $\mm$-core and the $\mm$-support are related in a simple manner:

\begin{remark}\label{PS4a}
    $c_{\mm}(Q) + s_{\mm}(I - Q) = I$ for all $Q \in \pr$.
\end{remark}

\begin{remark}\label{PS4b} 
    Core and support have the following properties: 
    \begin{displaymath}
c_{\mm}(\We_{\kik}P_{k}) = \We_{\kik}c_{\mm}(P_{k}), \ \ s_{\mm}(\Ve_{\kik}P_{k}) = \Ve_{\kik}s_{\mm}(P_{k})
\end{displaymath}
and 
\begin{displaymath}
c_{\mm}(\Ve_{\kik}P_{k}) \geq \Ve_{\kik}c_{\mm}(P_{k}), \ \ s_{\mm}(\We_{\kik}P_{k}) \leq \We_{\kik}s_{\mm}(P_{k}).
\end{displaymath}
\end{remark}

\begin{lemma}\label{PS5}
    Let $E = (\el)_{\gl \in \RR}$ be a spectral family in $\rr$ and
    for $\gl \in \RR$ define
    \[
	(c_{\mm}E)_{\gl} := c_{\mm}(\el), \quad (s_{\mm}E)_{\gl} :=
	\We_{\mu > \gl}s_{\mm}(\emm).
    \]
    Then $c_{\mm}E := ((c_{\mm}E)_{\gl})_{\gl \in \RR}$ and
    $s_{\mm}E := ((s_{\mm}E)_{\gl})_{\gl \in \RR}$ are spectral
    families in $\mm$.
\end{lemma}
\emph{Proof:} If $\gl < \mu$ then $c_{\mm}(\el) ≤ \el ≤ \emm$ and
therefore $c_{\mm}(\el) ≤ c_{\mm}(\emm)$. Moreover $\We_{\mu >
\gl}c_{\mm}(\emm) ≤ \We_{\mu > \gl}\emm = \el$, hence 
\[
    \We_{\mu > \gl}c_{\mm}(\emm) ≤ c_{\mm}(\el) ≤ \We_{\mu >
    \gl}c_{\mm}(\emm).
\]
The other assertions are obvious. Note, however, that $\gl \tto
s_{\mm}(\el)$ isn't a spectral family in general! \ \ $\Box$ \\

\begin{proposition}\label{PS6}
    Let $f \in \orr$ and let $E$ be the spectral family corresponding 
    to $f$. Then $c_{\mm}E$ is the spectral family corresponding to
    $\gr_{\mm}f$.
\end{proposition}
\emph{Proof:} Let $\kI$ be a dual ideal in $\pmm$. Then
$(\gr_{\mm}f)(\kI) = f(C_{\rr}(\kI))$ and
\begin{eqnarray*}
    f(C_{\rr}(\kI)) & = & \inf \{ \gl \ | \ \el \in C_{\rr}(\kI) \}  \\
     & = & \inf \{ \gl \ | \ \ex \ P \in \kI : \ P ≤ \el \}  \\
     & = & \inf \{ \gl \ | \ c_{\mm}(\el) \in \kI \}.
\end{eqnarray*}
Thus the assertion follows from the theorem that an observable function defines a unique spectral family (\cite{deg4}).  \ \ $\Box$ \\
~\\
By theorem 2.6 in \cite{deg4}, the restriction map $\gr_{\mm} : \orr \to
\omm$ induces a restriction map
\[
\begin{array}{cccc}
    \gr_{\mm} : & \hr & \to & \mm_{sa}  \\
     & A & \tto & \gr_{\mm}A
\end{array}
\]
for selfadjoint operators. In particular, we obtain

\begin{corollary}\label{PS7}
    $\gr_{\mm}Q = s_{\mm}(Q)$ for all projections $Q$ in $\rr$.
\end{corollary}
The corollary shows that the restriction map $\gr_{\mm} : \hr \to
\mm_{sa}$ has the important property that it maps projections to
projections and acts as the identity on $\pmm$. It also shows that
in general $\gr_{\mm}$ is not linear: if $P, Q \in \pr$ such that $PQ 
=0$ then it is possible that $s_{\mm}(P)s_{\mm}(Q) \ne 0$ and
therefore $s_{\mm}(P + Q) \ne s_{\mm}(P) + s_{\mm}(Q)$.\\
~\\
We will now consider the special case $\mm := Q\rr Q$. This case will 
show up the link to the restriction of ordinary continuous functions
$f : M \to \RR$ on a topological space $M$ to an open subspace $U
\tm M$. \\
~\\
Of course there is another natural way to restrict an operator to the 
subalgebra $Q\rr Q$, namely the map
\[
    \begin{array}{cccc}
	 & \rr & \to & Q\rr Q  \\
	 & A & \tto & QAQ.
    \end{array}
\]
But this type of restriction does not have the property that it maps
projections to projections:

\begin{remark}\label{PS7a}
    If $P \in \pr$, then $QPQ$ is a projection if and only if $P$ commutes
    with $Q$. 
\end{remark}
\emph{Proof:} If $QPQ$ is a projection, then
\[
    P \we Q = \lim_{n \to \∞}(PQ)^n = \lim_{n \to \∞}(QPQ)^n = QPQ,
\]
where the limits are taken with respect to the strong topology. Hence
\[
    (P - QPQ)(Q - QPQ) = PQ - QPQ = 0, 
\]
because $Q - QPQ = Q(I - P)Q$, so $Q(I - P)Q$ equals $(I - P) \we Q$
and, therefore, is a subprojection of $I - P$. It is then obvious that
$PQ = QP$ holds. \ \ $\Box$ \\

\begin{proposition}\label{PS8}
    Let $A \in \hr$ and let $\ea$ be the spectral family of $A$. Then 
    the spectral family of the restriction $\gr_{Q\rr Q}A$ is given by
    $\gl \tto \eal \we Q$.
\end{proposition}
\emph{Proof:} A projection $P \in \rr$ is an element of $Q\rr Q$ if
and only if $P ≤ Q$. Hence if $E \in \pr$ and $P \in \kP(Q\rr Q)$ such
that $P ≤ E$ then $P ≤ E \we Q$. This shows $c_{Q\rr Q}(E) = E \we Q$.
Therefore the proposition follows from proposition \ref{PS6}. \ \
$\Box$ \\ 
~\\
Note that for $\rr =\lh$ the subalgebra $Q\rr Q$ is canonically
isomorphic to $\kL(Q\kh)$, so $\gr_{Q\lh Q}A$ can be considered as the
restriction of $A \in \lh$ to an operator in $\kL(Q\kh)$. \\
Now let $f : M \to \RR$ be a continuous function on a topological
space $M$ and let $U \tm M$ be an open nonvoid subset. The
corresponding spectral family is given by $\gs_{f} : \gl \tto int
(\overset{-1}{f}(]-\∞, \gl]))$. Then $\gl \tto int (\overset{-1}{f}(]-\∞, \gl]))
\cap U$ is a spectral family in $\kT(U)$. Because of
\begin{eqnarray*}
    int (\overset{-1}{f}(]-\∞, \gl])) \cap U & = & int (\overset{-1}{f}(]-\∞, \gl])
    \cap U)  \\
     & = & int (\overset{-1}{(f_{| U})}(]-\∞, \gl]))
\end{eqnarray*}
this is the spectral family of the continuous function $f_{| U} : U
\tto \RR$, the restriction of $f$ to $U$. This also demonstrates that our
definition of restriction of operators is absolutely natural. \\
~\\
Proposition \ref{PS6} and lemma \ref{PS5} suggest still 
another natural possibility for defining a restriction map $\gs_{\mm} 
: \hr \to \mm_{sa}$: if $\ea$ is the spectral family corresponding to
$A \in \hr$ then $\gs_{\mm}A$ is the selfadjoint operator defined by
the spectral family $s_{\mm}\ea$. \\
Let us check what this means in the case $\mm = Q\rr Q$ for some $Q
\in \pr$ different from $I$. First of all we have to determine the
$Q\rr Q$-support of a projection $E \in \rr$. Here a little bit of
care is needed because $Q\rr Q$ is not a von Neumann subalgebra of
$\rr$ in the strict sense. It has a unity, $Q$, but this is different 
from $I$. This was irrelevant for the $Q\rr Q$-core but $\{ P \in Q\rr Q
\ | \ E ≤ P \} = \emptyset$ unless $E ≤ Q$. We can overcome this
complication by defining $\We_{Q\rr Q}\emptyset := Q$. Then we obtain
for all $E \in \pr$
\[
    s_{Q\rr Q}(E) = 
    \begin{cases}
	E   & \text{if} \ \ E \in Q\rr Q   \\
	Q   & \text{otherwise}
    \end{cases}
\]
and therefore the spectral family of $\gs_{Q\rr Q}A$ is given by
\[
    (s_{Q\rr Q}\ea)_{\gl} =
    \begin{cases}
	\eal   &  \text{if} \ \ \ex \ \mu > \gl : \ \ea_{\mu} \in Q\rr Q 
	\\
	Q       &  \text{otherwise}.   
    \end{cases}
\]
~\\
We will show that the restriction map $\gs_{\mm} : \hr \to \mm_{sa}$ has a canonical origin too.\\
~\\
Let $g: \por \to \RR$ be a mirrored observable function. 
There is a result analogous to proposition \ref{PS3} for the restriction $\gs_{\mm}g$ of $g$ to $\mm$: 

\begin{proposition}\label{PS8a} 
$\all \ \kI \in \kD (\mm) : \ (\gs_{\mm}g) (\kI) = g (C_{\rr}(\kI)). $
\end{proposition}
\emph{Proof:} $(\gs_{\mm}g) (\kI) = \sup \{ s (P) \mid P \in \kI \} = \sup \{ s (Q) \mid Q \in C_{\rr}(\kI) \}.$ \ \ $\Box$\\
~\\
Let $E$ be the spectral family of $A \in \hr$. It is now easy, to show that $\gs_{\mm}E := (\We_{\mu > \gl}s_{\mm}(\emm))_{\lir}$ is the spectral family of the operator $\gs_{\mm}A \in \mm_{sa}$ corresponding to $\gs_{\mm}g_{A}$. 

\begin{corollary}\label{PS8b} 
Let $A \in \hr$ and let $E$ be the spectral family of $A$. Then $\gs_{\mm}E$ is the spectral family corresponding to $\gs_{\mm}g_{A}$. 
\end{corollary}
\emph{Proof:} Indeed, we obtain for all dual ideals $\kI$ in $\pmm$: 
\begin{eqnarray*}
\gs_{\mm}g_{A}(\kI)  & = & g_{A}(C_{\rr}\kI)  \\
 & = & -f_{-A}(C_{\rr}\kI)  \\
 & = & -\inf \{ \lir \mid \ex \ P \in \kI : P \leq c_{\mm}(I - E_{-\gl-}) \} \\  
 & = & -\inf \{ \lir \mid c_{\mm}(I - E_{-\gl-}) \in \kI \} \\
 & = & -\inf \{ -\lir \mid c_{\mm}(I - E_{\gl-}) \in \kI \}  \\
 & = & \sup \{ \lir \mid c_{\mm}(I - E_{\gl}) \in \kI \}  \\
 & = & \sup \{ \lir \mid I - s_{\mm}(\el) \in \kI \} \\
 & = & \sup \{ \lir \mid I - \We_{\mu > \gl}s_{\mm}(\emm) \in \kI \},
\end{eqnarray*}  
where we have used some elementary properties of $\inf$ and $\sup$. \ \ $\Box $\\
~\\
There is a simple relation between the two types of
restrictions that is quite analogous to that between observable and
mirrored observable functions:

\begin{proposition}\label{PS8b}
    Let $\mm$ be a von Neumann subalgebra of $\rr$. Then
    \[
	\gs_{\mm}A = - \gr_{\mm}(- A)
    \]
    holds for all $A \in \hr$.
\end{proposition}
\emph{Proof:} Because of
\begin{eqnarray*}
    - f_{\gs_{\mm}A}(P) & = & - \inf \{ \lir \mid \We_{\mu >
    \gl}s_{\mm}(\eamu) ≥ P \}  \\
     & = & - \inf \{ \lir \mid s_{\mm}(\eal) ≥ P \}  \\
     & = & - \inf \{ \lir \mid I - c_{\mm}(I - \eal) ≥ P \}  \\
     & = & \sup \{ \lir \mid I - c_{\mm}(I - \ea_{- \gl}) ≥ P \} \\
     & = & g_{\gr_{\mm}(- A)}(P)
\end{eqnarray*}
for all $P \in \pom$, we obtain
\[
    f_{\gs_{\mm}A} = - g_{\gr_{\mm}(- A)} = - g_{- (- \gr_{\mm}(-
    A))} = f_{- \gr_{\mm}(- A)},
\]
and this implies
\[
    \gs_{\mm}A = - \gr_{\mm}(- A). \ \ \Box
\]
~\\

If $\mm$ is an arbitrary von Neumann subalgebra of $\rr$ and $A \in
\hr$ then
\[
    \all \ \gl \in \RR : \ (\gr_{\mm}\ea)_{\gl} ≤ \eal ≤
    (\gs_{\mm}\ea)_{\gl}
\]  
which means $\gs_{\mm}A ≤_{s} A ≤_{s} \gr_{\mm}A$ and therefore
$\gs_{\mm}A ≤ A ≤ \gr_{\mm}A$ by \cite{deg2}. \\

\begin{proposition}\label{PS8a}
    Let $\mm$ be a von Neumann subalgebra of the von Neumann algebra
    $\rr$. Then, for all $A \in \hr$, we have
    \[
	\gs_{\mm}A = \Ve \{ B \in \mm_{sa} \ | \ B ≤_{s} A \}
    \]
    and 
    \[
	\gr_{\mm}A = \We \{ C \in \mm_{sa} \ | \ A ≤_{s} C \}, 
    \]
    where $\gs_{\mm}A, \gr_{\mm}A$ are considered as elements of 
    $\rr$ and $\Ve, \We$ denote the greatest lower bound and the
    least upper bound with respect to the spectral order.
\end{proposition}
\emph{Proof:} Let $B, C \in \mm_{sa}$ such that $B ≤_{s} A ≤_{s} C$
and let $\ea, \eb, \ec$ be the spectral families of $A, B$ and $C$ 
respectively. Then, by the definition of the spectral order, we have
for all $\gl \irr$
\[
    \ecl ≤ \eal ≤ \ebl,
\]  
and therefore, using $\ebl, \ecl \in \mm$, we obtain
\begin{eqnarray*}
    \ecl & = & c_{\mm}(\ecl) \\
     & ≤ & c_{\mm}(\eal)   \\
     & ≤ & \eal  \\
     & ≤ & s_{\mm}(\eal)  \\
     & ≤ & \We_{\mu > \gl}s_{\mm}(\eamu)  \\
     & ≤ & \We_{\mu > \gl}s_{\mm}(\ebmu)  \\
     & = & \We_{\mu > \gl}\ebmu  \\
     & = & \ebl.
\end{eqnarray*}
This shows
\[
    B ≤_{s} \gs_{\mm}A ≤_{s} A ≤_{s} \gr_{\mm}A ≤_{s} C. \ \ \Box
\]
~\\
This proposition shows that the two restriction mappings $\gr_{\mm}$
and $\gs_{\mm}$ from $\hr$ onto $\mm_{sa}$ are on an equal footing. Moreover, it shows that these restrictions are generalisations of $\mm$- support and $\mm$- core to arbitrary selfadjoint operators. { \bf We call $\gr_{\mm}A$ the \emph{upper $\mm$- aspect of $A$} and $\gs_{\mm}A$ the \emph{lower $\mm$- aspect of $A$}}.\\
~\\
Let $m_{A} := \inf \{ \gl \ | \ \eal \ne 0 \}$ and $M_{A} := \min \{
\gl \ | \ \eal =I \}$. Then
\[
    m_{A}I = \gs_{\CC I}A \ \ \text{and} \ \ M_{A}I = \gr_{\CC I}A.
\]  
Thus we recover via restrictions the well known simple inequality
\[
    m_{A}I ≤ A ≤ M_{A}I.
\]
In general, $\gs_{\mm}A, \gr_{\mm}A$ can be considered as \emph{lower 
and upper}, respectively, \emph{coarse grainings} of $A$. The
following example makes this point of view apparent.

\begin{example}\label{PS8b}
    Let $A \in \hr$ and let $\gl_{1}, \ldots, \gl_{n} \in sp(A)$ such 
    that $\gl_{1} < \cdots < \gl_{n}$. We may assume that the
    corresponding spectral projections satisfy $\ea_{\gl_{1}} < \cdots
    < \ea_{\gl_{n}}$. Let
    \[
	\kaa := \kaa(\gl_{1}, \ldots , \gl_{n})
    \]
    be the von Neumann subalgebra generated by $\{\ea_{\gl_{1}}, \ldots ,
    \ea_{\gl_{n}}, I\}$, that is $\kaa = lin_{\CC}\{\ea_{\gl_{1}}, \ldots ,
    \ea_{\gl_{n}}, I\}$. Setting $\ea_{\gl_{0}} := 0$ and $\ea_{\gl_{n + 1}}
    := I$, we can represent every projection $P \in \kaa$ as a linear combination 
    \[
	P = \sum_{k = 1}^{n + 1}a_{k}(\ea_{\gl_{k}} - \ea_{\gl_{k -
	1}})
    \]  
    with coefficients $a_{k} \in \{0, 1\}$. In order to avoid boring
    case distinctions, we further assume that
    \[
	m_{A} < \gl_{1} < \gl_{n} < M_{A}.
    \]
    We can therefore set $\gl_{n + 1} := M_{A}$.
    Then an easy, but somewhat tedious, discussion shows that for all
    $\gl \irr$ such that $\eal \ne 0$ we have
    \[
	c_{\kaa}(\eal) =
	\begin{cases}
	0     & \quad \text{if} \quad \eal < \ea_{\gl_{1}}      \\
	\ea_{\gl_{k}}  & \quad \text{if} \quad \ea_{\gl_{k}} ≤ \eal <
	\ea_{\gl_{k + 1}} \quad (k = 1,\ldots, n)  \\
	I     & \quad \text{if} \quad \eal = I   
	\end{cases}
    \]
    and
    \[
	s_{\kaa}(\ea)_{\gl} =
	\begin{cases}
	\ea_{\gl_{1}}     & \quad \text{if} \quad \eal < \ea_{\gl_{1}}      \\
	\ea_{\gl_{k + 1}}  & \quad \text{if} \quad \ea_{\gl_{k}} ≤ \eal <
	\ea_{\gl_{k + 1}} \quad (k = 1,\ldots, n - 1)  \\
	I     & \quad \text{if} \quad \ea_{\gl_{n}} ≤ \eal   
	\end{cases}.
    \]
    Therefore, the spectra of the restrictions $\gr_{\kaa}A$ and
    $\gs_{\kaa}A$ are
    \[
	sp(\gr_{\kaa}A) = \{ \gl_{1}, \ldots, \gl_{n}, M_{A} \} 
    \]
    and
    \[
	sp(\gs_{\kaa}A) = \{ m_{A}, \gl_{1}, \ldots, \gl_{n} \}.
    \]
    It follows that the restrictions $\gr_{\kaa}A$ and $\gs_{\kaa}A$
    have spectral representations
    \[
	\gr_{\kaa}A = \sum_{k = 1}^{n + 1}\gl_{k}(\ea_{\gl_{k}} -
	\ea_{\gl_{k - 1}})
    \]
    and
    \[
	\gs_{\kaa}A = m_{A}\ea_{\gl_{1}} + 
	\sum_{k = 1}^{n}\gl_{k}(\ea_{\gl_{k + 1}} - \ea_{\gl_{k}})
    \]
    respectively. These are finite approximations of the spectral
    representation $A = \int_{\RR}\gl d\eal$ of $A$: $\gr_{\kaa}A$ is 
    the upper and $\gs_{\kaa}A$ is the lower Riemann-Stieltjes sum
    defined by the partition $(m_{A}, \gl_{1}, \ldots, \gl_{n}, M_{A})$. 
\end{example}

\section{The upper and lower observable presheaves} 
\label{ulp} 

Consider three von Neumann subalgebras $\kaa, \kbb, \kcc$ of $\rr$
such that $\kaa \tm \kbb \tm \kcc$. Then the corresponding restriction
maps $\kcb : \kcc_{sa} \to \kbb_{sa}, \ \kba : \kbb_{sa} \to \kaa_{sa}$
and $\kca : \kcc_{sa} \to \kaa_{sa}$ obviously satisfy
\begin{equation}
    \kca = \kba \circ \kcb \quad \text{and} \quad \gr^{\kaa}_{\kaa} =
    id_{\kaa_{sa}}.
    \label{eq:ps}
\end{equation}
The set $\frsr$ of all von Neumann subalgebras of $\rr$ is a lattice
with respect to the partial order given by inclusion. The meet of
$\kaa , \kbb \in \frsr$ is defined as the intersection,
\[
    \kaa \we \kbb := \kaa \cap \kbb,
\]
and the join as the subalgebra generated by $\kaa$ and $\kbb$:
\[
    \kaa \vee \kbb := (\kaa \cup \kbb)''.
\]
The join is a rather intricate operation. This can already be seen in
the most simple (non-trivial) example $lin_{\CC}\{I, P\} \vee
lin_{\CC}\{I, Q\}$ for two non-commuting projections $P, Q \in \rr$
(see \cite{kr4}). Fortunately we don't need it really. \\

The subset $\frAr \tm \frsr$ of all \emph{abelian} von Neumann
subalgebras of $\rr$ is also partially ordered by inclusion but it is 
only a \emph{semilattice}: the meet of two (in fact of an arbitrary
family of) elements of $\frAr$ always exists but the join does not in 
general. Both $\frsr$ and $\frAr$ have a smallest element, namely
$\mathbb{O} := \CC I$. However, unless $\rr$ is itself abelian, there
is no greatest element in $\frAr$. Anyway, $\frsr$ and $\frAr$ can be 
considered as the sets of objects of (small) \emph{categories} whose
morphisms are the inclusion maps. \\
In quantum physics the (maximal) abelian von Neumann subalgebras of
$\lh$ are called \emph{contexts}. We generalize this notion in the
following

\begin{definition}\label{PS9}
    The small category $\cor$, whose objects are the abelian von Neumann
    subalgebras of $\rr$ and whose morphisms are the inclusion maps,
    is called the {\bf context category of the von Neumann algebra
    $\rr$}.
\end{definition}
Since the morphisms of $\cor$ are so simple, we also speak of $\frAr$
as the context category or the \emph{category} of abelian von Neumann
subalgebras of $\rr$. \\        

We define a \emph{presheaf} $\kO^{+}_{\rr}$ on the context category $\cor$
of $\rr$ by sending objects $\kaa \in \frAr$ to $\kO^{+}_{\rr}(\kaa) :=
\kaa_{sa}$ (or equivalently to $\oaa$) and morphisms $\kaa
\hookrightarrow \kbb$ to restrictions $\kba : \kbb \to \kaa$. Due to
\ref{eq:ps} this gives a contravariant functor, i.e. a presheaf on
$\cor$.
\begin{definition}\label{PS10}
    The presheaf $\kO^{+}_{\rr}$ is called the {\bf upper observable presheaf} of
    the von Neumann algebra $\rr$.
\end{definition}

\begin{remark}\label{PS10a}
     We can define a presheaf $\kO^{-}_{\rr}$ by using the restrictions
     $\gs^\kbb_{\kaa}: \kbb_{sa} \to \kaa_{sa}, \ A \tto \gs_{\kaa}A$,
     for $\kaa \hookrightarrow \kbb$ in $\cor$. This presheaf is called the lower 
     observable presheaf. Due to the next result, it has quite analogous
     properties as $\kO^{+}_{\rr}$, so we will concentrate on $\kO^{+}_{\rr}$.
\end{remark}

\begin{proposition}
The presheaves $\kO^{+}_{\rr}$ and $\kO^{-}_{\rr}$ are isomorphic. 
\end{proposition}
\emph{Proof:} This is a direct consequence of the fact that $- \orr$ is the set of mirrored observable functions. The most simple way to describe the isomorphism is to regard observable functions as completely increasing functions. Then, for all $\kaa \in \frAr$, 
\begin{displaymath}
\begin{array}{cccc}   
  \gF_{\kaa}: & \kO (\kaa)  & \to & - \kO (\kaa)     \\
      & f & \tto & - f 
\end{array}
\end{displaymath}
is obviously a bijection that commutes with restrictions: for all $\kaa,  \kbb \in \frAr$ such that $\kaa \tm \kbb$ we have 
\begin{displaymath}
\gF_{\kbb}(f) _{|\poa} = \gF_{\kaa}(f_{|\poa})
\end{displaymath} 
for all $f \in \kO (\kbb)$. Hence $\gF:= (\gF_{\kaa})_{\kaa \in \frAr}$ is an isomorphism from $\kO^{+}_{\rr}$ onto $\kO^{-}_{\rr}$.  \ \ $\Box$
 
Each observable function $f \in \orr$ induces a family
$(f_{\kaa})_{\kaa \in \frAr}$, defined by $f_{\kaa} := \gr_{\kaa}f$,
which is compatible in the following sense:
\[
    \all \ \kaa, \kbb \in \frAr : \ \gr^{\kaa}_{\kaa \cap \kbb}f_{\kaa}
    = \gr^{\kbb}_{\kaa \cap \kbb}f_{\kbb}.
\]
The problem whether each compatible family $(f_{\kaa})_{\kaa \in \frAr}$
is induced by an observable function in $\orr$ will be discussed in
section \ref{gs}. \\ 

We will now show how the restriction maps $\kba$ and $\gs^\kbb_{\kaa}$
act on observable functions $f : \qb \to \RR$ or, in other words, how the
Gelfand transformation behaves with respect to the restrictions $\kba$
and $\gs^\kbb_{\kaa}$.   

\begin{lemma}\label{PS11}
    Let $\kaa, \kbb$ be abelian von Neumann algebras such that $\kaa
    \tm \kbb$. Then $\pi^{\kbb}_{\kaa} : \gb \tto \gb \cap \kaa$ maps 
    $\qb$ onto $\qa$. The mapping $\pi^{\kbb}_{\kaa}$ is continuous,
    open and therefore also identifying. Moreover
    \[
	\gb \cap \kaa = \{ s_{\kaa}(P) \ | \ P \in \gb \}.
    \]
\end{lemma}
\emph{Proof:} This lemma is a special case of proposition 2.1 in \cite{deg6}. 
$\Box$ \\

\begin{proposition}\label{PS12} 
    Let $\kaa$ be an abelian von Neumann algebra and $\kj$ a dual
    ideal in $\pa$. Then 
    \[
        \kj = \bigcap \{ \gb \in \qa \mid \kj \tm \gb \}. 
    \]
\end{proposition}
\emph{Proof:} Let $\kQ_{\kj}(\kaa) := \{ \gb \in \qa \mid \kj \tm \gb
\}$. Assume that there is $E \in \bigcap \kQ_{\kj}(\kaa)$ such that $E
\notin \kj$. Since $\kj$ is a dual ideal, this implies $P(I - E) = P - PE \ne 0$
for all $P \in \kj$. The commutativity of $\kaa$ implies that $(I - E)\kj$ 
is a filter base. Since $(I - E)P ≤ P, I - E$, the cone $C_{\kaa}((I - E)\kj)$ is a
dual ideal in $\pa$ that contains $\kj$ and $I - E$. But then any
quasipoint that contains $C_{\kaa}((I - E)\kj)$, contains $\kj$, hence
also $E$ and, by construction, $I - E$, a contradiction. \ \ $\Box$\\
~\\
It is obvious that the proposition is true for all Boolean algebras
$\kbb$: every dual ideal in $\kbb$ is the intersection of
quasipoints.

\begin{corollary}\label{PS12a}
    Let $\kaa$ be an abelian von Neumann subalgebra of the abelian von
    Neumann algebra $\kbb$ and let $\gga \in \qa$. Then
    \[
	C_{\kbb}(\gga) = \bigcap \{ \gb \in \qb \ | \ \gga \tm \gb \}.
    \]
\end{corollary}

 \begin{corollary}\label{PS13}
    Let $\kaa, \kbb$ be as above and let $f : \qb \to \RR$ be an observable
    function. Then we have for all $\gga \in \qa$: 
    \begin{enumerate}
	\item  [(i)] $(\kba f)(\gga) = \sup \{ f(\gb) \ | \ \gga \tm \gb
	\}$ and            
    
	\item  [(ii)] $(\gs^\kbb_{\kaa}f)(\gga) = \inf \{ f(\gb) \ | \
	\gga \tm \gb \}$.      
    \end{enumerate}
\end{corollary}
\emph{Proof:} $(i)$ This follows immediately from propositions \ref{PS3},
\ref{PS12} and general properties of observable functions:
\[
    (\kba f)(\gga) = f(C_{\kbb}(\gga)) = f(\bigcap \{ \gb \in \qb \ | \ 
    \gga \tm \gb \}) = \sup \{ f(\gb) \ | \ \gga \tm \gb \}. 
\]
$(ii)$ \ Since $f$ is the Gelfand transform of some $A \in \kbb_{sa}$ and $f=f_{A}=g_{A}$ on $\qb$, we obtain from proposition \ref {PS8a} and from theorem \ref {in2} 
\begin{displaymath}
(\gs^\kbb_{\kaa}f)(\gga) = g_{A}(C_{\kbb}(\gga)) = g_{A}(\bigcap_{\gga \tm \gb}\gb) = \inf_{\gga \tm \gb}g_{A}(\gb) = \inf_{\gga \tm \gb}f(\gb). \ \ \Box
\end{displaymath}
~\\
Let $\pi := \pi^\kbb_{\kaa} : \qb \to \qa$ be the identifying mapping 
$\gb \tto \gb \cap \pa$ from $\qb$ onto $\qa$. By proposition 2.1 in 
\cite{deg6}, $\pi$ is continuous, open and satisfies
\[
    \all \ P \in \pb : \ \pi(\qpb) = \kQ_{s_{\kaa}(P)}(\kaa).
\] 
The fibres $\urb{\pi}(\gga), \ \gga \in \qa,$ form a partition of
$\qb$ into closed subsets. Typically, they have empty interior. Since 
$\qb$ is a Stonean space, there is a unique $P \in \pb$ such that
\[
     int \urb{\pi}(\gga) = \qpb.  
\]
Hence, if $int \urb{\pi}(\gga) \ne \emptyset$,
\[
    \{\gga\} = \pi(int \urb{\pi}(\gga)) = \pi(\qpb) = \kQ_{s_{\kaa}(P)}
    (\kaa)
\]
and therefore $\{\gga\}$ is an open closed set. This means that $\gga$
is an \emph{atomic quasipoint} of $\pa$. In this case, moreover,
$\urb{\pi}(\gga)$ is open and closed, and this implies $P \in \pa$. So
we have proved:

\begin{remark}\label{PS13a}
    For every $\gga \in \qa$, the fibre $\urb{\pi}(\gga)$ of $\pi :=
    \pi^\kbb_{\kaa}$ is open and closed if and only if $\gga$ is an
    atomic quasipoint. If $\gga$ is not atomic, then the interior of
    $\urb{\pi}(\gga)$ is empty. 
\end{remark}

If $\kaa$ has finite dimension, then every quasipoint in $\pa$ is
atomic, so $\{ \urb{\pi}(\gga) \ | \ \gga \in \qa \}$ is a finite
partition of $\qb$ into open closed subsets.  \\

\section{A unification of upper and lower observable presheaves}	
\label{uni} 

We define a presheaf $\kO_{\rr}$ on $\frAr$ such that $\kO^{+}_{\rr}$ and $\kO^{-}_{\rr}$ are subpresheaves of $\kO_{\rr}$. We note first that the sum of two completely increasing functions $f, h : \por \to \RR$ is, in general, not completely increasing: 

\begin{example}
Let $P, Q \in \por$ such that $PQ = 0$ and let $f := r_{P}, \ h := r_{Q}, \ E := I - P, \ F := I - Q$. Then 
\begin{displaymath}
f (E) = 0, \ f (F) = 1, \ h (E) = 1, \ h (F) = 0,
\end{displaymath}    
so 
\begin{displaymath}
(f + h) (E) = (f +h) (F) = 1,
\end{displaymath}
but 
\begin{displaymath}
f (E \vee F) = h (E \vee F) = 1,
\end{displaymath}
hence 
\begin{displaymath}
(f + h) (E \vee F) = 2 > 1 = \max ( (f + h) (E), (f + h) (F)). 
\end{displaymath}
\end{example}
  
We have seen that the presheaves $\kO^{+}_{\rr}$ and $\kO^{-}_{\rr}$ can be defined by the ordinary restriction of observable and mirrored observable functions, respectively. This restriction is a {\bf linear} (and multiplicative) operation. The sets $\orr$ and $-\orr$, however, have only poor algebraic structure: they are partially ordered and closed with respect to multiplication by nonnegative real numbers. Therefore, we are led to introduce the real vector space $F_{\rr}$ generated by the set $F^{+}_{\rr}$ of completely increasing functions. This space is generated equally well by the set $F^{-}_{\rr}$ of completely decreasing functions. Note that we can construct $F_{\rr}$ also in the following way: Let $M^{+}_{\rr}$ be the additive monoid generated by $F^{+}_{\rr}$. Since $F^{+}_{\rr}$ is closed under multiplication by nonnegative real numbers, $M^{+}_{\rr}$ is a cone in the space of functions on $\por$. Similarly, the additive monoid $M^{-}_{\rr}$ generated by $F^{-}_{\rr}$ is a cone and can be represented as $M^{-}_{\rr} = - M^{+}_{\rr}$. Eventually,  we have 
\begin{displaymath}
F_{\rr} = M^{+}_{\rr} + M^{-}_{\rr}. 
\end{displaymath}
Moreover, it is easy to see that $F_{\rr}$ is isomorphic to the Grothendieck group (\cite{la}) of the monoid $M^{+}_{\rr}$. 

\begin{definition}
We define a presheaf $\kO_{\rr}$ on $\frAr$ by 
\begin{enumerate}
\item [ (i)] $\kO_{\rr}(\kaa) := F_{\kaa}$ for all $\kaa \in \frAr$ and 
\item [ (ii)] $\gr^{\kbb}_{\kaa}(f) := f_{\mid \poa}$ for $\kaa \tm \kbb, \ f \in F_{\kbb}$.   
\end{enumerate}
$\kO_{\rr}$ is called the { \bf observable presheaf of $\rr$}. 
\end{definition}
Note that the presheaves $\kO^{+}_{\rr}$ and $\kO^{-}_{\rr}$ can be embedded in $\kO_{\rr}$ as subpresheaves. This follows immediately from the fact that, by construction, the presheaf $\kO^{+}_{\rr}$ is isomorphic to the presheaf of completely increasing functions on the context category $\cor$ of $\rr$ and, analogously, that the presheaf $\kO^{-}_{\rr}$ is isomorphic to the presheaf of completely decreasing functions on $\cor$. \\
In contrast with the upper and lower observable presheaves, $\kO_{\rr}$ is a presheaf of real vector spaces and linear maps. But it is not at all obvious how to interpret it at the level of operators or even physically. We leave this problem to future work.

\section{Global Sections}
\label{gs}

Let $\rr$ be a von Neumann algebra acting on a Hilbert space $\kh$.
We will now consider the upper and lower observable presheaves $\kO^{\pm}_{\rr}$ on the context category $\cor$, defined in the previous section, more
closely. We restrict our considerations to the upper presheaf because for the lower presheaf results and proofs are completely analogous. \\

We have seen that every observable function $f \in \orr$ induces a
family $(f_{\kaa})_{\kaa \in \frAr}$ of observable functions $f_{\kaa}
\in \oaa$, defined by $f_{\kaa} := \gr^{\rr}_{\kaa}f$. This family has
the following compatibility property:
\begin{equation}
    \all \ \kaa, \kbb \in \frAr : \ \gr^{\kaa}_{\kaa \cap \kbb}f_{\kaa}
	= \gr^{\kbb}_{\kaa \cap \kbb}f_{\kbb}.
    \label{eq:gs1}
\end{equation}
$(f_{\kaa})_{\kaa \in \frAr}$ is therefore a \emph{global section} of 
the presheaf $\kO^{+}_{\rr}$ in the following general sense.

\begin{definition}\label{gs1}
    Let ${\bf C}$ be a category and $\kS : {\bf C} \to {\bf Set}$ a presheaf,
    i.e. a contravariant functor from ${\bf C}$ to the category ${\bf
    Set}$ of sets. A global section of $\kS$ assigns to every object 
    $a$ of ${\bf C}$ an element $\gs(a)$ of the set $\kS(a)$ such that
    for every morphism $\gf : b \to a$ of ${\bf C}$
    \[
	\gs(b) = \kS(\gf)(\gs(a))
    \]
    holds.
\end{definition}

Not every presheaf admits global sections. An important example is the
{\bf spectral presheaf} of the von Neumann algebra $\rr$. This is the 
presheaf $\kS : \frAr \to {\bf CO}$ from the category $\frAr$ of
abelian von Neumann subalgebras of $\rr$ to the category ${\bf CO}$ of
compact Hausdorff spaces which is defined by
\begin{enumerate}
    \item  [(i)] $\kS(\kaa) := \qa$ for all $\kaa \in \frAr$,

    \item  [(ii)] $\kS(\kaa \hookrightarrow \kbb) :=
    \pi^{\kbb}_{\kaa}$, where the mapping $\pi^{\kbb}_{\kaa} : \qb
    \to \qa$ is defined in lemma \ref{PS11}.  
\end{enumerate}

We know from theorem 3.2 in \cite{deg3} that there is a canonical
homeomorphism $\go_{\kaa} : \qa \to \gO(\kaa)$. This homeomorphism
intertwines the ordinary restriction
\[
\begin{array}{cccc}
    r^\kbb_{\kaa} : & \gO(\kbb) & \to & \gO(\kaa)  \\
     & \gt & \tto & \gt |_{\kaa}
\end{array}
\]
with $\pi^{\kbb}_{\kaa}$:
\[
    r^\kbb_{\kaa} \circ \go_{\kbb} = \go_{\kaa} \circ \pi^\kbb_{\kaa}.
\]
This shows, according to a reformulation of the Kochen-Specker theorem
by J. Hamilton, C.J. Isham and J. Butterfield (\cite{ish3}, \cite{doe})
that the presheaf $\kS : \frAr \to {\bf CO}$ admits no global
sections.\\

In the case of the observable presheaf $\kO^{+}_{\rr}$ there are plenty of
global sections because each $A \in \hr$ induces one. Here the natural
question arises whether all global sections of $\kO^{+}_{\rr}$ are induced
by selfadjoint elements of $\rr$. This is certainly not true if the
Hilbert space $\kh$ has dimension two. For in this case the
constraints \ref{eq:gs1} are void and therefore \emph{any} function on
the complex projective line defines a global section of $\kO^{+}_{\lh}$.
But Gleason's (or Kochen-Specker's) theorem teaches us that the
dimension two is something peculiar. We will show, however, that the
phenomenon, that there are global sections of $\kO_{\rr}$ that are not 
induced by selfadjoint elements of $\rr$, is not restricted to
dimension two.

\begin{definition}\label{gs2}
    We denote by $\gG(\kO^{+}_{\rr})$ the set of global sections of the
    observable presheaf $\kO^{+}_{\rr}$. The image of the canonical
    mapping
    \[
	\begin{array}{cccc}
	    \gS_{\rr} : & \orr & \to & \gG(\kO^{+}_{\rr})  \\
	     & f & \tto & (\gr^\rr_{\kaa}f)_{\kaa \in \frAr}
	\end{array}
    \]
    is denoted by $\gS(\orr)$.
\end{definition}

We will show in the sequel that $\gG(\kO^{+}_{\rr})$ is strictly larger
than $\gS(\orr)$ for $\rr = \kL(\CC ^3)$. The example that we shall
give for this case can be generalized easily to higher dimensions.\\

We begin with a general

\begin{remark}\label{gs3}
    Let $f \in \orr$ such that $\gS_{\rr}(f)$ is a family of observable
    functions of projections. Then also $f$ is the observable function
    of a projection.
\end{remark}
\emph{Proof:} Let $A \in \hr$ such that $f = f_{A}$. Then $A \in \kaa$
for some $\kaa \in \frAr$ and therefore $im f = im \gr^\rr_{\kaa}f
\tm \{0, 1\}$. Hence $A$ is a projection. \ \ $\Box$ \\

Now we consider the special case $\rr := \lh, \kh := \CC^3$ more
closely.

\begin{remark}\label{gs4}
    Let $E \in \rr$ be a projection of rank two and $f := f_{E}$ 
    the corresponding observable function. If $P \in \puh$, i.e. a projection
    of rank one, then 
    \[
	f(P) = \begin{cases}
	0  &  \ \text{if} \ P = I - E  \\
	1  &  \ \text{otherwise}.
	\end{cases}
    \]
    Hence $f(P) = 0$ for \emph{exactly one} $P \in \puh$, namely $P = I - 
    E$.  
\end{remark}

Each twodimensional von Neumann subalgebra $\kaa$ of $\kL(\CC^3)$ is
abelian and generated by exactly one projection $P \in \puc$. Moreover
the maximal abelian von Neumann subalgebras $\mm$ of $\kL(\CC^3)$ are
threedimensional and they are determined by orthogonal triples $(P_{1},
P_{2}, P_{3}) \in \puc^3$: $\mm = lin_{\CC}\{P_{1}, P_{2}, P_{3}\}$.
Two such triples determine the same algebra $\mm$ if and only if one
is a permutation of the other.\\

Now let $P_{1}, P_{2} \in \puc$ be projections that do not commute
(i.e. $P_{1}P_{2} \ne 0$) and for $k = 1, 2$ let
\[
    \frA_{k} := \{ \kaa \in \frA(\kL(\CC^3)) \ | \ P_{k} \in \kaa \}.  
\]
Define a family $(Q_{\kaa})_{\kaa \in \frA(\kL(\CC^3))}$ of projections
$Q_{\kaa} \in \kaa$ by
\[
    Q_{\kaa} := 
    \begin{cases}
	I - P_{k}  & \ \text{if} \ \kaa \in \frA_{k} \quad (k = 1, 2) 
	\\
	I             & \ \text{if} \ \kaa \notin \frA_{1} \cup
	\frA_{2} 
    \end{cases}
\]
and let $f_{\kaa}$ be the observable function of $Q_{\kaa}$. We show
that $(f_{\kaa})_{\kaa \in \frA(\kL(\CC^3))}$ is a global section of
$\kO_{\kL(\CC^3)}$. In order to do that it is convenient to work
directly with the projections $Q_{\kaa}$ instead with their observable
functions because restricting a projection $P$ to $\kaa \in
\frA(\kL(\CC^3))$ means passing to its $\kaa$-support $s_{\kaa}(P)$.\\
We have to prove that the constraints \ref{eq:gs1} are satisfied for
the family $(Q_{\kaa})_{\kaa \in \frA(\kL(\CC^3))}$. Since $I$ remains
unchanged by restriction we have to control the behaviour of $I -
P_{k}, \ (k = 1, 2)$. Now
\[
    (I - P_{k})|_{\kaa} =
    \begin{cases}
	I - P_{k} & \ \text{if} \ P_{k} \in \kaa  \\
	I            &  \ \text{otherwise},
    \end{cases}
\]
where $|_{\kaa}$ is a shortcut for the restriction to $\kaa$. If $\kaa
\in \frA_{k}$, $\kbb \notin \frA_{k}$ and $\kcc \tm \kaa \cap \kbb$,
then $P_{k} \notin \kcc$ and therefore $(I - P_{k})|_{\kcc} =I$. If
$\kaa \in \frA(\kL(\CC^3))$ has dimension two and is contained in
$\kaa_{1} \cap \kaa_{2}$, where $\kaa_{k} \in \frA_{k} \ (k = 1, 2)$, 
then $\kaa = lin_{\CC}\{Q, I - Q\}$ with $Q \in \puc$ orthogonal to
$P_{1} \vee P_{2}$. But then $P_{1}, P_{2} \notin \kaa$ and therefore 
$(I - P_{k})|_{\kaa} = I$ for $k = 1, 2$. Altogether this shows that
$(f_{\kaa})_{\kaa \in \frA(\kL(\CC^3))}$ is a global section. \\
This global section cannot be induced by an observable function $f \in
\kO(\kL(\CC^3))$: If this were the case then, by remark \ref{gs3},
$f$ would be the observable function of a projection $E$ and,
according to the definition of the family $(f_{\kaa})_{\kaa \in
\frA(\kL(\CC^3))}$, $E$ must be of rank two. But then
\[
    f(P_{1}) = f(P_{2}) = 0, 
\]
a contradiction to remark \ref{gs4}. Therefore we have proved:

\begin{proposition}\label{gs5}
    $\gG(\kO_{\kL(\CC^3)})$ is strictly larger than
    $\gS(\kO(\kL(\CC^3)))$.
\end{proposition}

This leads us to the following

\begin{definition}\label{gs6}
    Let $\rr$ be a von Neumann algebra. The global sections of the observable
    presheaf $\kO^{+}_{\rr}$ are called {\bf (upper) contextual observables}.
\end{definition}
Clearly, $\gG(\kO^{+}_{\rr}) = \gS(\orr)$ if $\rr$ is abelian. \\
~\\
Contextual observables can be characterized as certain functions on
$\por$:

\begin{proposition}\label{gs7}
    Let $\rr$ be a von Neumann algebra. There is a one-to-one
    correspondence between global sections of the observable presheaf
    $\kO^{+}_{\rr}$ and functions $f : \por \to \RR$ that satisfy
    \begin{enumerate}
	\item  [(i)] $f(\Ve_{k \ikk}P_{k}) = \sup_{k \ikk}f(P_{k})$ for all
	{\bf commuting} families $(P_{k})_{k \ikk}$ in $\por$,
    
	\item  [(ii)] $f_{|_{\por \cap \kaa}}$ is bounded for all
	$\kaa \in \frAr$.
    \end{enumerate}
\end{proposition}
\emph{Proof:} Let $(f_{\kaa})_{\kaa \in \frAr}$ be a global section of 
$\kO^{+}_{\rr}$. Then the functions $f_{\kaa} : \por \cap \kaa \ (\kaa
\in \frAr)$ can be glued to a function $f : \por \to \RR$:\\
Let $P \in \por$ and let $\kaa$ be an abelian von Neumann subalgebra
of $\rr$ that contains $P$. Then 
\[
    f(P) := f_{\kaa}(P)
\]
does not depend on the choice of $\kaa$. Indeed, if $P \in \kaa \cap
\kbb$, then $f_{\kaa}(P) = f_{\kbb}(P)$ by the compatibility property 
of global sections. It is obvious that $f$ satisfies properties $(i)$ 
and $(ii)$. \\
If, conversely, a function $f : \por \to \RR$ with the properties
$(i)$ and $(ii)$ is given and if $\kaa$ is an abelian von Neumann subalgebra
of $\rr$, then $f_{\kaa} := f_{|_{\por \cap \kaa}}$ is a completely
increasing function. The family $(f_{\kaa})_{\kaa \in \frAr}$ is then,
by construction, a global section of $\kO^{+}_{\rr}$. \ \ $\Box$ \\

Note that the conditions in proposition \ref{gs7} are much weaker than
the condition of being completely increasing, so it is not surprising 
that there are global sections of $\kO_{\rr}$ which are not induced by
a single selfadjoint operator.

\section{States}
\label{st}

Although ``states'' do not belong to our proper theme, we will include some remarks about quantum states emphasising again the presheaf perspective. \\
~\\
A \emph{state of a von Neumann algebra $\rr$} is a positive linear
functional $\gf : \rr \to \CC$ with $\gf(I) = 1$. We denote the
(convex and weak* compact) set of states of $\rr$ by $\kS(\rr)$.\\
~\\
There is a natural restriction of states to von Neumann subalgebras of $\rr$:

\begin{definition}\label{st1}
    Let $\mm$ be a von Neumann subalgebra of $\rr$. The usual restriction
    of mappings defines a restriction map
    \[
	\begin{array}{cccc}
	    st^\rr_{\mm} : & \ksr & \to & \ksm  \\
	     & \gf & \tto & \gf|_{\mm}.
	\end{array} 
    \]
\end{definition}
$st^\rr_{\mm}$ is a \emph{surjective} mapping (see \cite{kr1}, p.266)
and for any three von Neumann subalgebras $\kaa, \kbb, \kcc$ of $\rr$ 
such that $\kaa \tm \kbb \tm \kcc$ we have the obvious properties
\begin{equation}
    st^\kcc_{\kaa} = st^\kbb_{\kaa} \circ st^\kcc_{\kbb} \quad
	\text{and} \quad st^\kaa_{\kaa} = id_{\ksa}.
    \label{eq:st1}
\end{equation}
If we consider in particular the \emph{abelian} von Neumann subalgebras
of $\rr$ we obtain, due to \ref{eq:st1}, a presheaf $\kS_{\rr}$ that is in
some sense dual to the observable presheaf $\kO_{\rr}$:

\begin{definition}\label{st2}
    The contravariant functor $\kS_{\rr} : \frAr \to {\bf Set}$,
    defined on objects by
    \[
	\kS_{\rr}(\kaa) := \kS(\kaa)
    \]
    and on morphisms by 
    \[
	\kS_{\rr}(\kaa \hookrightarrow \kbb) := st^\kbb_{\kaa},
    \]
    is called the {\bf state presheaf} of the von Neumann algebra
    $\rr$.
\end{definition}
Each state $\gf \in \kS(\rr)$ gives rise to a global section
$(\gf_{\kaa})_{\kaa \in \frAr}$, defined by
\[
    \gf_{\kaa} := \gf |_{\kaa},
\]
of the presheaf $\kS_{\rr}$. Also here arises the natural question
whether all global sections of $\kS_{\rr}$ are of this form. \\

We will show in the sequel that the answer is affirmative, provided
that $\rr$ has no direct summand of type $I_{2}$. So here again the
dimension two forms the notorious exception. Nevertheless this
contrasts to the situation of the observable presheaf where we can
find counterexamples in all dimensions.\\

Let $(\gf_{\kaa})_{\kaa \in \frAr}$ be a global section of $\kS_{\rr}$.
If $A \in \hr$ then $A \in \kaa$ for some $\kaa \in \frAr$. If $A$
belongs also to $\kbb \in \frAr$ then $A \in \kaa \cap \kbb$ and
therefore
\[
    \gf_{\kaa}(A) = \gf_{\kaa \cap \kbb}(A) = \gf_{\kbb}(A).
\]
This shows that the global section $(\gf_{\kaa})_{\kaa \in \frAr}$
determines a function $\gf : \hr \to \RR$, defined by
\[
    \all \ A \in \hr \ \all \ \kaa \in \frAr : \ (A \in \kaa \ \lra \ \gf(A) =
    \gf_{\kaa}(A)).  
\]
Clearly $\gf(I) =1$ and $\gf(A^\ast A) ≥ 0$ for all $A \in \rr$. The
salient point is the $\RR$-linearity of $\gf$.\\

Indeed, if $\rr = \kL(\CC^2)$, then $\gf$ may fail to be linear. To see
this, note that the maximal abelian von Neumann subalgebras of
$\kL(\CC^2)$ are of the form $\kaa_{P} := lin_{\CC}\{P, I - P\}$ with a
projection $P \ne 0, I$. A state $\gf_{\kaa_{P}}$ on $\kaa_{P}$ is
given by prescribing an arbitrary value $a_{P} \in [0, 1]$ to $P$ and 
the condition $\gf_{\kaa_{P}}(I - P) = 1 - a_{P}$. Because the
intersection of two different maximal abelian von Neumann subalgebras 
of $\kL(\CC^2)$ equals $\CC I$, there are no constraints for a global 
section of the state presheaf. Now assume that each global section of 
the state presheaf $\kS_{\kL(\CC^2)}$ is induced by a state of 
$\kL(\CC^2)$. Let $P, Q \in \kL(\CC^2)$ be two noncommuting
projections of rank one. Since $\kL(\CC^2)$ is noncommutative, $P + Q$
generates a maximal abelian von Neumann subalgebra $\kaa_{R}$ of
$\kL(\CC^2)$ and, because $P$ does not commute with $Q$, the
subalgebras $\kaa_{P}, \kaa_{Q}$ and $\kaa_{R}$ are pairwise different.
Now
\begin{equation}
    P + Q = aR + b(I - R)
    \label{eq:st2}
\end{equation}
with uniquely determined real numbers $a, b$. To each choice of
$a_{P}, a_{Q}, a_{R} \in [0, 1]$ there is a global section and
therefore by assumption a state $\gf$ of $\kL(\CC^2)$ such that
\[
    \gf(P) = a_{P}, \gf(Q) =a_{Q}, \gf(R) = a_{R}. 
\]
Choosing $a_{P} = a_{Q} = a_{R} = 0$, 
\ref{eq:st2} implies $b = 0$ and the choice $a_{P} = a_{Q} = 0, a_{R} 
= 1$ leads to $a = 0$. This contradicts \ref{eq:st2}.\\

We return to the discussion of the mapping $\gf : \hr \to \RR$ defined
by a global section $(\gf_{\kaa})_{\kaa \in \frAr}$ of $\kS_{\rr}$.
Let $P_{1}, \ldots, P_{n} \in \pr$ be pairwise orthogonal. Then
$P_{1}, \ldots, P_{n} \in \kaa$ for some $\kaa \in \frAr$ and
therefore
\[
    \gf(\sum_{j = 1}^{n}P_{j}) = \gf_{\kaa}(\sum_{j = 1}^{n}P_{j}) =
    \sum_{j = 1}^{n}\gf_{\kaa}(P_{j}) = \sum_{j = 1}^{n}\gf(P_{j}).
\]
This implies that $\gf|_{\pr} : \pr \to [0, 1]$ is a (finitely additive)
\emph{probability measure on the projection lattice $\pr$}. Here we
can apply a substantial generalization of Gleason's theorem, due to
Christensen, Yeadon et al.:

\begin{theorem}\label{st3}(\cite{mae}, thm. 12.1)
    Let $\rr$ be a von Neumann algebra without direct summand of type 
    $I_{2}$ and let $\mu : \pr \to [0, 1]$ be a finitely additive
    probability measure on $\pr$. Then $\mu$ can be extended to a
    unique state of $\rr$.        
\end{theorem}

It follows from the spectral theorem that a state $\gf$ of $\rr$ is
uniquely determined by its restriction to $\pr$. Hence we obtain from 
theorem \ref{st3} and the previous discussion:

\begin{theorem}\label{st4}
    The states of a von Neumann algebra $\rr$ without direct summand
    of type $I_{2}$ are in one to one correspondence to the global
    sections of the state presheaf $\kS_{\rr}$. This correspondence is
    given by the bijective map
    \[
	\begin{array}{cccc}
	    \gG_{\rr} : & \kS(\rr) & \to & \gG(\kS_{\rr})  \\
	     & \gf & \tto & (\gf |_{\kaa})_{\kaa \in \frAr}.
	\end{array}
    \]
\end{theorem}
Clearly, the core of this theorem is the surjectivity of the mapping
$\gG_{\rr}$. We will show that this property implies that each
probability measure on $\pr$ can be extended to a state of $\rr$. Thus
theorem \ref{st4} is indeed equivalent to theorem \ref{st3}.

\begin{proposition}\label{st5}
    The following properties of a von Neumann algebra $\rr$ are
    equivalent:
    \begin{enumerate}
	\item  [(i)] Every finitely additive measure on $\pr$ extends 
	to a state of $\rr$.
    
	\item  [(ii)] Every global section of $\kS_{\rr}$ is induced
	by a state of $\rr$.
    \end{enumerate}
\end{proposition}
\emph{Proof:} According to the foregoing discussion it remains to show
that a probability measure $\mu$ on $\pr$ defines a global section of 
$\kS_{\rr}$. Let $\kaa \in \frAr$. Then $\mu_{\kaa} := \mu |_{\pa}$ is
a probability measure on $\pa$. We construct from $\mu_{\kaa}$ a state
$\gf_{\kaa}$ of $\kaa$ that extends $\mu_{\kaa}$. This construction is
quite similar to that used in the proof that the Gelfand spectrum of $\kaa$ is homeomorphic to its Stone spectrum (\cite {deg3}).  
Let
\begin{equation}
    A := \sum_{j = 1}^{m}a_{j}P_{j}
    \label{eq:st3}
\end{equation}
where $P_{1}, \ldots, P_{m} \in \pa$ are pairwise orthogonal and
$a_{1}, \ldots, a_{m}$ are complex numbers. Since we do not assume
that the coefficients $a_{j}$ are all different from zero, we can and 
do assume that $\sum_{j = 1}^{m}P_{j} = 1$, i.e. that $(P_{1}, \ldots,
P_{m})$ is a partition of unity. We then call \ref{eq:st3} a \emph{normalized
representation} of $A \in lin_{\CC}\pa$. Each element of
$lin_{\CC}\pa$ has a normalized representation. In order to extend 
$\mu_{\kaa}$ linearly we are forced to define
\begin{equation}
    \gf_{\kaa}(A) := \sum_{j = 1}^{m}a_{j}\mu_{\kaa}(P_{j})
    \label{eq:st4}
\end{equation}
for $A \in lin_{\CC}\pa$ given by \ref{eq:st3}. To show that this is
well defined, consider another normalized representation $\sum_{k =
1}^{n}b_{k}Q_{k}$ of $A$. If $x \in im(P_{j}Q_{k})$ then $a_{j}x = Ax 
= b_{k}x$, hence
\begin{equation}
    \all \ j, k \ : \ (P_{j}Q_{k} \ne 0 \ \lra \ a_{j} = b_{k}).
    \label{eq:st5}
\end{equation}
This implies
\[
    \sum_{j}a_{j}\mu_{\kaa}(P_{j}) = \sum_{j,
    k}a_{j}\mu_{\kaa}(P_{j}Q_{k}) = \sum_{j, k}b_{k}\mu_{\kaa}(P_{j}Q_{k})
    = \sum_{k}b_{k}\mu_{\kaa}(Q_{k}).
\]
Clearly $\gf_{\kaa}(aA) = a\gf_{\kaa}(A)$ for all $a \in \CC$ and $A
\in lin_{\CC}\pa$ in normalized representation. If $A, B \in
lin_{\CC}\pa$ have normalized representations $\sum_{j}a_{j}P_{j},
\sum_{k}b_{k}Q_{k}$ respectively then
\begin{eqnarray*}
    A + B & = & \sum_{j}a_{j}P_{j} + \sum_{k}b_{k}Q_{k}  \\
     & = & \sum_{j, k}a_{j}P_{j}Q_{k} + \sum_{j, k}b_{k}P_{j}Q_{k}  \\
     & = & \sum_{j, k}(a_{j} + b_{k})P_{j}Q_{k}
\end{eqnarray*} 
and therefore
\begin{eqnarray*}
    \gf_{\kaa}(A + B) & = & \sum_{j, k}(a_{j} + b_{k})\mu_{\kaa}
    (P_{j}Q_{k})  \\
     & = & \gf_{\kaa}(A) + \gf_{\kaa}(B).
\end{eqnarray*}
It is obvious that $A \in lin_{\CC}\pa$ is positive if in a normalized
representation $\sum_{j}a_{j}P_{j}$ of $A$ all coefficients are
nonnegative. Hence $\gf_{\kaa} : lin_{\CC}\pa \to \CC$ is a positive
linear functional with $\gf_{\kaa}(I) = \mu_{\kaa}(I) = 1$. Now
\[
    |\sum_{j}a_{j}P_{j}| = \max_{j}|a_{j}|
\]
for every normalized representation $\sum_{j}a_{j}P_{j}$ of $A \in
lin_{\CC}\pa$ and 
\begin{eqnarray*}
    |\gf_{\kaa}(\sum_{j}a_{j}P_{j})| & = &
    |\sum_{j}a_{j}\mu_{\kaa}(P_{j})|  \\
     & ≤ & \sum_{j}|a_{j}|\mu_{\kaa}(P_{j})  \\
     & ≤ & \max_{j}|a_{j}|,               
\end{eqnarray*}
so $|\gf_{\kaa}| = 1$. This implies that $\gf_{\kaa}$ has a unique
extension to $\kaa = \overline{lin_{\CC}\pa}$ and that this extension,
which we also denote by $\gf_{\kaa}$, is a state of $\kaa$.\\
If $\kaa, \kbb \in \frAr$ such that $\kaa \tm \kbb$ then trivially
$\mu_{\kaa} = \mu_{\kbb} |_{\pa}$ and therefore, according to the
foregoing construction, $\gf_{\kaa} = \gf_{\kbb} |_{\kaa}$. So we have
constructed a global section $(\gf_{\kaa})_{\kaa \in \frAr}$ from the 
probability measure $\mu$ on $\pr$. \ \ $\Box$ \\

Each state $\gf_{\kaa}$ of $\kaa \in \frAr$ can be seen as a positive
Radon measure on $C(\qa)$: 
\[
    \begin{array}{cccc}
	\gfa : & C(\qa) & \to & \CC  \\
	 & \psi & \tto & \gfa(A_{\psi})
    \end{array}
\]
where $A_{\psi} \in \kaa$ is obtained from $\psi \in C(\qa)$ by the
inverse of the Gelfand transformation. Therefore it induces a probability
measure $\nua$ on $\qa$ such that
\begin{equation}
    \all \ \psi \in C(\qa) : \ \gfa(\psi) = \int_{\qa}\psi d\nua.
    \label{eq:st6}
\end{equation}
Identifying the Gelfand transform of $A \in \kaa$ with the
(complexified) observable function $f^{A}_{\kaa}$ we can write this as
\begin{equation}
    \gfa(A) = \int_{\qa}f^{A}_{\kaa}d\nua.
    \label{eq:st7}
\end{equation}
From the definition of $\nua$ we get immediately
\begin{equation}
     \all \ P \in \pa : \ \gfa(P) = \nua(\qpa).
    \label{eq:st8}
\end{equation}
We denote by $\mm^1(\qa)$ the set of probability measures on $\qa$.
These sets form a presheaf $\mm^1_{\rr}$ on the category $\frAr$. For
$\kaa, \kbb \in \frAr$ such that $\kaa \tm \kbb$ we define 
\[
    \begin{array}{cccc}
	p^\kbb_{\kaa} : & \mm^1(\qb) & \to & \mm^1(\qa)  \\
	 & \nu & \tto & \pi^\kbb_{\kaa}\nu,
    \end{array}
\]
where $\pi^\kbb_{\kaa}\nu$ denotes the \emph{image of the measure
$\nu$} under the continuous mapping $\pi^\kbb_{\kaa} : \qb \to \qa, \
\gb \tto \gb \cap \pa$. It is defined by
\begin{equation}
    \all \ \psi \in C(\qa) : \ \int_{\qa}\psi d\pi^\kbb_{\kaa}\nu :=
    \int_{\qb}(\psi \circ \pi^\kbb_{\kaa})d\nu
    \label{eq:st9}
\end{equation}              
or, equivalently, by
\begin{equation}
    (\pi^\kbb_{\kaa}\nu)(M) := \nu(\overset{-1}{\pi^\kbb_{\kaa}}(M))
    \label{eq:st10}
\end{equation}          
for all Borel subsets $M$ of $\qa$.
It is obvious that for $\kaa, \kbb, \kcc \in \frAr$ such that $\kaa
\tm \kbb \tm \kcc$ we have
\begin{equation}
    p^\kcc_{\kaa} = p^\kbb_{\kaa} \circ p^\kcc_{\kbb} \ \text{and} \
    p^\kaa_{\kaa} = id_{\mm^1(\qa)}.
    \label{eq:st11}
\end{equation}
So $\mm^1_{\rr}$, together with the restriction mappings
$p^\kbb_{\kaa}$, is a presheaf on $\frAr$. \\

We recall the following well known

\begin{definition}\label{st6}
    Let $\rr$ be a von Neumann algebra. A state $\gf$ of $\rr$ is
    called normal if 
    \[
	\gf(\sup_{k \ikk}A_{k}) = \sup_{k \ikk}\gf(A_{k})
    \]
    for every increasing bounded net $(A_{k})_{k \ikk}$ in $\rr$.\\
    If $\rr$ is abelian, $\rr = C(\qr)$, then a probability measure $\nu$
    on $\qr$ is called normal if the corresponding state $\gf^\nu :
    \psi \tto \int_{\qr}\psi d\nu$ of $\rr$ is normal. 
\end{definition}

We call a global section of $\kS_{\rr}$ or $\mm^1_{\rr}$ normal if all
of its members are normal.

\begin{lemma}\label{st7}
    Let $\gf$ be a state of a von Neumann algebra $\rr$. Then $\gf$ is
    normal if and only if $\gf_{\kaa} := \gf |_{\kaa}$ is normal for
    all $\kaa \in \frAr$.
\end{lemma}
\emph{Proof:} It is obvious that all $\gf_{\kaa} \ (\kaa \in \frAr)$
are normal if $\gf$ is normal. For the converse we use the result that
a state $\gf$ is normal if it is normal on $\pr$, i.e. if $\gf(\sum_{k
\ikk}P_{k}) = \sum_{k \ikk}\gf(P_{k})$ for every orthogonal family
$(P_{k})_{k \ikk}$ of projections in $\rr$ (\cite{kr2}, Thm. 7.1.12). 
Because every such family is contained in a suitable $\kaa \in \frAr$,
the assertion follows. \ \ $\Box$
		      
\begin{proposition}\label{st8}
    The presheaves $\mm^1_{\rr}$ and $\kS_{\rr}$ over $\frAr$ are
    isomorphic. This isomorphism maps normal sections to normal
    sections.
\end{proposition}
\emph{Proof:} The isomorphism is simply given by the family
$(\gF_{\kaa})_{\kaa \in \frAr}$ of (convex-linear) maps
\[
    \begin{array}{cccc}
	\gF_{\kaa} : & \mm^1(\qa) & \to & \kS(\kaa)  \\
	 & \nu & \tto & \gf^\nu
    \end{array}
\]
where $\gf^\nu$ is defined by $\gf^\nu(A) := \int_{\qa}
\kf_{\kaa}(A)d\nu$. The Riesz representation theorem ensures that the 
mappings $\gF_{\kaa}$ are convex-linear isomorphisms. So it remains to
show that the $\gF_{\kaa}$ are compatible with restrictions:
\[
    \all \ \kaa, \kbb \in \frAr : \ (\kaa \tm \kbb \ \lra \
    (\gF_{\kbb}(\nu_{\kbb})) |_{\kaa} =
    \gF_{\kaa}(p^\kbb_{\kaa}(\nu_{\kbb}))).   
\]
Due to linearity and continuity it suffices to check this for
projections. If $P \in \pa$ then, using \ref{eq:st8}, we have
\begin{eqnarray*}
    (\gF_{\kbb}(\nub))(P) & = & \nub(\qpb)  \\
     & = & \nub(\overset{-1}{\pi^\kbb_{\kaa}}(\qpa))  \\
     & = & (p^\kbb_{\kaa}(\nub))(\qpa)  \\
     & = & (\gF_{\kaa}(p^\kbb_{\kaa}(\nub))(P).
\end{eqnarray*}
The normality assertion is obvious. \ \ $\Box$ \\

Let $\kh$ be a separable Hilbert space of dimension greater than two
and $\gf$ a normal state of $\lh$. The theorem of \emph{Gleason}
(\cite{kr2}) assures that $\gf$ is induced by a positive traceclass
operator $\gr$ of trace one:
\[
    \all \ A \in \lh : \ \gf(A) = tr (\gr A).
\] 
If $\gr = P_{\CC x}$ with $x \in S^1(\kh)$, then $\gf(A) = <Ax, x>$,
and in this case $\gf$ is called a \emph{vector state}. We recall that a normal state of a maximal abelian von Neumann subalgebra of a von Neumann algebra is always the restriction of a vector state (\cite{kr2}). The following result describes the situation when this restriction is a pure state.  

\begin{theorem}\label{st9}
Let $\kh$ be a separable Hilbert space of dimension greater than two,
$\mm \tm \lh$ a maximal abelian von Neumann subalgebra of $\lh$ and
$\gf$ a normal state of $\lh$. Then the Radon measure $\mu$ on
$\kQ(\mm)$, induced by $\gf_{|_{\mm}}$, is the point measure
$\eps_{\gb_{0}}$ for some $\gb_{0} \in \kQ(\mm)$, if and only if
there is an $x \in S^1(\kh)$ such that $P_{\CC x} \in \mm, \ \gb_{0} =
\gb_{\CC x}$ and $\gf$ is the vector state defined by $x$.
\end{theorem}
\emph{Proof:} Let $x \in S^1(\kh)$ such that $P_{\CC x} \in \mm$ and 
let $\rho := P_{\CC x}$. Let $\psi$ be a real valued continuous function on 
$\kQ(\mm)$ and let $A_{\psi} \in \mm$ be the corresponding 
hermitian operator.
$P_{\CC x}$ commutes with $A_{\psi}$, so $x$ is an eigenvector of 
$A_{\psi}$. Let $\gl$ be the corresponding eigenvalue. Then
\begin{displaymath}
    tr(\rho A_{\psi}) = <A_{\psi}x, x> = \gl<x, x> = \gl.
\end{displaymath}
On the other hand
\begin{displaymath}
    \gl  = f_{A_{\psi}}(\gb_{\CC x}) = 
    \eps_{\gb_{\CC x}}(f_{A_{\psi}}) = \eps_{\gb_{\CC x}}(\psi).
\end{displaymath}
Therefore
\begin{displaymath}
    \mu = \eps_{\gb_{\CC x}}.
\end{displaymath}

Conversely, let $\gf$ be a normal state, $\gf = tr (\gr -)$, and let
$\mm$ be a maximal abelian von Neumann subalgebra of $\lh$ such 
that the measure $\mu$ on $\qmm$, corresponding to $\gf_{|_{\mm}}$
is the point measure $\eps_{\gb_{0}}$ for some $\gb_{0} \in \kQ(\mm)$.
Then for all $P \in \pmm$
\begin{displaymath}
    tr(\rho P) = \mu(\chi_{\kQ_{P}(\mm)}) = 
    \chi_{\kQ_{P}(\mm)}(\gb_{0})
\end{displaymath}
and therefore
\begin{displaymath}
    \forall \ P \in \pmm : \ (P \in \gb_{0} \ \Longleftrightarrow \ 
    tr(\rho P) = 1).
\end{displaymath}
Let $P \in \gb_{0}$ and let $(e_{k})_{k \in \NN}$ be an 
$P$-adapted orthonormal basis of $\kh$, i.e. $e_{k} \in im P \cup 
ker P$ for all $k \in \NN$. Then
\begin{eqnarray*}
    1 & = & tr(\rho P)  \\
     & = & tr(P \rho)  \\
     & = & \sum_{k}<P\rho e_{k}, e_{k}>  \\
     & = & \sum_{k}<\rho e_{k}, Pe_{k}>  \\
     & = & \sum_{e_{k} \in im P}<\rho e_{k}, e_{k}>.
\end{eqnarray*}
Because of $<\rho e_{k}, e_{k}> \geq 0$ for all $k \in \NN$ and 
$tr\rho = 1$ we conclude that
\begin{displaymath}
    \forall \ e_{k} \in ker P : \ <\rho e_{k}, e_{k}> = 0.
\end{displaymath}
Hence $\rho e_{k} = 0$ for all $e_{k} \in ker P$ and therefore
\begin{displaymath}
    \rho(I - P) = 0
\end{displaymath}
i.e.
\begin{displaymath}
    \rho = \rho P.
\end{displaymath}
In particular
\begin{displaymath}
    \rho P = \rho = \rho^{*} = P \rho.
\end{displaymath}
This implies that $\gr$ commutes with $\mm$ and therefore, since $\mm$
is maximal abelian, $\gr \in \mm$. Hence the range projection
$P_{\overline{im \gr}}$ belongs to $\pmm$. Now for all $y \in \kh$
\begin{displaymath}
    \rho y = \rho Py =P\rho y
\end{displaymath}
and therefore
\begin{displaymath}
    \forall \ P \in \gb_{0} \ \forall \ y \in \kh : \ \rho y \in im P.
\end{displaymath}
This implies
\begin{displaymath}
    \forall \ P \in \gb_{0} : \ \overline{im \rho} \tm im P,
\end{displaymath}
and from the maximality of $\gb_{0}$ we conclude that 
$P_{\overline{im \rho}} \in \gb_{0}$. $P_{\overline{im \rho}}$ is
therefore the minimal element of $\gb_{0}$. Let $P_{1} \in \lh$ be a
projection of rank one such that $P_{1} ≤ P_{\overline{im \rho}}$.
Then $P_{1}$ commutes with $\gb_{0}$ and therefore with $\pmm$, since 
$Q \in \gb_{0}$ or $I - Q \in \gb_{0}$ for all $Q \in \pmm$. Hence
$\overline{im \rho} = im \rho = \CC x$ for a unique line $\CC x$ in $\kh$.
Therefore $P_{\CC x} \in \pmm$ and $\gb_{0} = \gb_{\CC x}$.\\
There is a unique $\gl_{0} \in \CC$ such that
\begin{displaymath}
    \rho x = \gl_{0}x.
\end{displaymath}
$\rho \geq 0$ and $tr\rho = 1$ imply $\gl_{0} = 1$. Hence for all 
$y \in \kh$
\begin{displaymath}
    \rho^2 y = \rho(\rho y) = \rho(\gl x) = \gl\rho x = \gl x = \rho y
\end{displaymath}
and therefore
\begin{displaymath}
    \rho = P_{\CC x}. \ \ \Box 
\end{displaymath}
~\\
In classical physics, observables are continuous functions on some set of {\it pure states}. The connection with the definition of states of a von Neumann algebra is simply established by considering a pure state of classical physics as a point measure (evaluation functional) on a space of continuous functions. This is motivated by the definition of {\it mixed states} as probability measures (positive Radon measures of norm one) on the set of pure states. It should be discussed whether the linearity of states is due to this process of {\it averaging}. One advantage of the linearity of states is that the notion of superposition of states makes no (mathematical) difficulties.\\
~\\
We already mentioned in earlier parts of this work that, since observables can be considered as continuous functions on the Stone spectrum, one might think of the elements of the Stone spectrum as ``pure quasistates''. This, however, is a bit naive. It is obvious that a vector state $\pcx$ can be identified with the atomic quasipoint $\frb_{x}$, but $f_{A}(\frb_{x})$ does give the expectation value of $A$, if the physical system is in the state $\pcx$, only when $x$ is a normed eigenvector of $A$. We obtain the expectation value $<Ax, x>$ as 
\begin{displaymath}
<Ax, x> = f_{\pcx A \pcx}(\frb_{x})
\end{displaymath}
and we can prove (\cite{deg7}) that 
\begin{displaymath}
f_{\pcx A \pcx}(\frb_{x}) = \min_{P \in \frb_{x}}f_{PAP}(\frb_{x}). 
\end{displaymath}
This leads to a possible generalization for arbitrary quasipoints of an arbitrary von Neumann algebra: 

\begin{definition}
    Let $\rr$ be a von Neumann algebra, $A \in \hr$, $\frb \in \qr$
    and
    \[  
    \hat{\frb}(A) := \inf_{P \in \frb}f_{PAP}(\frb).
    \]
    The function 
    \[
    \begin{array}{cccc}
        \hat{\frb} : & \hr & \to & \RR \\
         & A & \tto & \hat{\frb}(A)
     \end{array}
     \]
     is called the quasistate of $\rr$ induced by $\frb$.
\end{definition}
Of course the same definition also applies to an arbitrary dual ideal 
in $\pr$.

\begin{remark}
    This definition is in accordance with the description of vector
    states, but also with the notion of pure state for an abelian von 
    Neumann algebra $\rr$: If $\gb \in \qr$ and $A \in \hr$, then
    \[
    \hat{\gb}(A) = \inf_{P \in \gb}f_{PAP}(\gb) = f_{A}(\gb)
    \]
    since $f_{PAP} = f_{P}f_{A}$ in the abelian case, and $f_{P}(\gb) =
    1$ for $P \in \gb$.
\end{remark}
The foregoing definition of an quasistate has the disadvantage that it is difficult to handle: If $P, Q \in \pr$ such that $P \leq Q$ does not imply that $f_{PAP} \leq f_{QAQ}$. For this would mean that $PAP \leq_{s} QAQ$, but even $PAP \leq QAQ$ is not true in general. Moreover, according to theorem of Neumark, it is hopeless to obtain the spectral family of $A$ from the spectral family of $PAP$ in a manageable way. Possibly, one can circumvent these difficulties by describing the observable function of $A$ in terms of the operator $A$, without using its spectral family.

\end{document}